\documentclass[twocolumn]{aastex63}

\usepackage{epsfig}
\usepackage{amsmath}
\usepackage{enumerate}
\usepackage{mathrsfs}
\usepackage{mathtools}
\usepackage{natbib}
\usepackage{units}
\usepackage{hyperref}
\usepackage{lineno}
\shorttitle{Axisymmetric Pulsar Magnetosphere}
\shortauthors{Hu \& Beloborodov}

\def\beq{\begin{equation}}
\def\eeq{\end{equation}}

\def\bmu{\boldsymbol{\mu}}
\def\bOm{\boldsymbol{\Omega}}
\def\bj{\boldsymbol{J}}
\def\bJ{\boldsymbol{J}}
\def\bB{\boldsymbol{B}}
\def\bE{\boldsymbol{E}}

\def\RLC{R_{\rm LC}}
\def\gthr{\gamma_{\rm thr}}
\def\Ish{I_{\rm sh}}
\def\dsh{\Delta_{\rm sh}}
\def\rhoco{\rho_{\rm co}}
\def\be{\boldsymbol{e}}
\def\br{\boldsymbol{r}}
\def\bS{\boldsymbol{S}}
\def\Ldiss{L_{\rm diss}}
\def\epph{\epsilon_{\rm ph}}
\def\Laz{\mathscr{L}}
\def\laz{l}

\def\Eq{Equation}

\newbox\grsign \setbox\grsign=\hbox{$>$} \newdimen\grdimen \grdimen=\ht\grsign
\newbox\simlessbox \newbox\simgreatbox \newbox\simpropbox
\setbox\simgreatbox=\hbox{\raise.5ex\hbox{$>$}\llap
     {\lower.5ex\hbox{$\sim$}}}\ht1=\grdimen\dp1=0pt
\setbox\simlessbox=\hbox{\raise.5ex\hbox{$<$}\llap
     {\lower.5ex\hbox{$\sim$}}}\ht2=\grdimen\dp2=0pt
\setbox\simpropbox=\hbox{\raise.5ex\hbox{$\propto$}\llap
     {\lower.5ex\hbox{$\sim$}}}\ht2=\grdimen\dp2=0pt

\def\simlt{\mathrel{\copy\simlessbox}}

\begin{document}

\title{Axisymmetric pulsar magnetosphere revisited}

\author{Rui Hu}
\affiliation{Physics Department and Columbia Astrophysics Laboratory, Columbia
  University, 538 West 120th Street New York, NY 10027}

\author{Andrei M. Beloborodov}
\affiliation{Physics Department and Columbia Astrophysics Laboratory, Columbia
  University, 538 West 120th Street New York, NY 10027}
\affiliation{Max Planck Institute for Astrophysics, Karl-Schwarzschild-Str. 1, D-85741, Garching, Germany}

\begin{abstract}
We present a global  kinetic plasma  simulation of an axisymmetric pulsar magnetosphere with self-consistent $e^\pm$ pair production. We use the particle-in-cell method and log-spherical coordinates with a grid size $4096\times 4096$. This allows us to achieve a high voltage induced by the pulsar rotation and investigate pair creation in a young pulsar far from the death line. We find the following. (1) The energy release and $e^\pm$ creation are strongly concentrated in the thin, Y-shaped current sheet, with a peak localized in a small volume at the Y-point. (2) The Y-point is shifted inward from the light cylinder by $\sim 15\%$, and ``breathes'' with a small amplitude. (3) The dense $e^\pm$ cloud at the Y-point is in ultra-relativistic rotation, which we call super-rotation, because it exceeds co-rotation with the star. The cloud receives angular momentum flowing from the star along the poloidal magnetic field lines. (4) Gamma-ray emission peaks at the Y-point and is collimated in the  azimuthal direction, tangent to the Y-point circle. (5) The separatrix current sheet between the closed magnetosphere and the open magnetic field lines is sustained by the electron backflow from the Y-point cloud. Its thickness is self-regulated to marginal charge starvation. (6) Only a small fraction of dissipation occurs in the separatrix inward of the Y-point. A much higher power is released in the equatorial plane, including
the Y-point where the created dense $e^\pm$ plasma is spun up and intermittently ejected through the nozzle between the two open magnetic fluxes.
\end{abstract}



\section{Introduction}

In order to understand the observed activity of pulsars one needs to solve a simply formulated physics problem: find the electromagnetic field and its dissipation rate around a rotating magnetized  sphere. The sphere can be treated as an ideal conductor, so that the magnetic field is ``frozen'' inside the rotating star.\footnote{The typical timescale for the magnetic field decay in neutron stars is millions of years, many orders of magnitude longer than relevant electrodynamic timescales outside the star.} A canonical version of this problem assumes that the star has a dipole magnetic field with the dipole moment $\boldsymbol{\mu}$ along the star's angular velocity $\boldsymbol{\Omega}$ --- the so-called ``aligned rotator.''

The problem would be simple if there were no plasma around the star \citep{pacini_energy_1967,pacini_rotating_1968, ostriker_nature_1969}, which would also imply no dissipation. It was quickly realized that this vacuum electromagnetic configuration is unrealistic \citep{goldreich_pulsar_1969}, because rotation induces an electric field $\bE$ outside the star with a component parallel to the magnetic field $\bB$. This field component $E_\parallel$ can extract particles from the stellar surface. Furthermore, if $E_\parallel$ is left unscreened, it accelerates a seed electron to an enormous Lorentz factor
\beq
\label{eq:gam0}
\gamma_0 = \frac{e\Phi_0}{m_ec^2}=\frac{\mu\Omega ^2}{m_e c^4}, 
\eeq 
where $m_e$ is the electron mass and $c$ is the speed of light. Particle acceleration triggers an avalanche of $e^\pm$ pair creation, and the magnetosphere becomes populated with plasma capable of conducting electric currents and screening $E_\parallel$. The plasma, is however, continually lost. Therefore, the new plasma-filled state must self-organize so that it sustains quasi-steady pair creation, which requires ongoing particle acceleration in some regions of the magnetosphere. This self-organization is a challenging nonlinear problem. It determines the energy release rate and radiation emitted by the pulsar.

\subsection{FFE model}

The pulsar picture significantly simplifies if one gives up following the plasma behavior and instead makes two assumptions: (1) the plasma everywhere screens $E_\parallel$ to zero, and (2) the plasma inertia is negligible. This description is often called force-free electrodynamics (FFE). It satisfies the condition $\rho \bE+\bj\times\bB/c=0$, so that there is no energy or momentum sink from the electromagnetic field. This condition determines the electric current density $\bj$, and thus closes the set of Maxwell equations for $\bE$ and $\bB$.

A steady FFE solution for the aligned rotator $\bmu\parallel\bOm$ was obtained numerically \citep{contopoulos_axisymmetric_1999, timokhin_force-free_2006}.
It demonstrates that the magnetosphere has a large ``closed zone'' with the approximately dipole magnetic field, and open twisted magnetic field lines, which extend from the polar caps of the star to the light cylinder, 
\beq 
  \RLC=\frac{c}{\Omega},
\eeq
in agreement with the picture described by \citet{goldreich_pulsar_1969}. The boundary of the closed zone intersects the equatorial plane on a circle of radius $R_Y$ (the ``Y-point’’). The minimum-energy FFE configuration has $R_Y=\RLC$.\footnote{When $R_Y=\RLC$, the magnetic field diverges at the Y-point while its volume-integrated energy remains finite \citep{Uzdensky03,Gruzinov05}. The divergence is absent if $R_Y<\RLC$.} 
The FFE magnetosphere was observed to relax to this configuration in time-dependent simulations
\citep{spitkovsky_time-dependent_2006, komissarov_simulations_2006, parfrey_introducing_2012}.  

At $r>R_Y$ an equatorial current sheet separates the two opposite magnetic fluxes that open to infinity. At $r<R_Y$ the current sheet is split into two symmetric parts (``separatrices'') that envelop the closed zone in the northern and southern hemispheres. The closed zone has no toroidal magnetic field, $B_\phi=0$, and so Stokes theorem requires zero net current through each polar cap. This is achieved with the separatrix current sheet, which screens the closed zone from the twisted field $B_\phi\neq 0$ in the open zone. The zero net current  also ensures that the charge of the star does not grow in the steady state. 
The equal positive and negative parts of the poloidal current are 
\beq 
\label{eq:I0}
  I_0\approx \frac{\mu\Omega^2}{c}.
\eeq 
The FFE condition implies that the poloidal current flows along the open poloidal magnetic field lines and can be measured at any sphere of radius $r\geq R_\star$ --- the current is independent of $r$, as expected in a steady state. The open field lines also carry a non-zero radial Poynting flux $S_r=cE_\theta B_\phi/4\pi$, and the net flux of electromagnetic energy from the rotating star (the ``spindown power'') is
\beq 
  L_{\rm sd}\approx\frac{\mu^2\Omega^4}{c^3}=I_0\Phi_0.
\eeq 

The FFE configuration demonstrates that the charge density sustained in the magnetosphere almost everywhere equals the co-rotation density \citep{goldreich_pulsar_1969},
\beq 
\rho_{\rm co}\approx -\frac{\bOm\cdot \bB}{2\pi c}.
\eeq 
The important exception is the Y-shaped current sheet, which carries a finite surface charge $\Sigma$. Thus, both $\bJ$ and $\rho$ diverge in the current sheet. The surface charge $\Sigma$ is determined by the jump conditions across the current sheet \citep{lyubarsky_equilibrium_1990}.

This FFE picture gives a first approximation to the electromagnetic field of a pulsar. It is based on neglecting dissipation, and calculations of the actual dissipation rate require going beyond FFE. A complete model must follow the plasma behavior in the magnetosphere, the  acceleration of particles, and $e^\pm$ creation. The observed pulsar activity cannot be understood without following these processes. 

The FFE solution offers hints about possible locations of particle acceleration, because it provides an overall picture of electric current density $\bj$ and charge density $\rho$ in the rotating magnetosphere. A useful parameter is the ratio 
\beq 
\label{eq:alpha}
  \alpha=\frac{J_\parallel}{c\rho}, 
\eeq 
where $J_\parallel$ is the electric current component parallel to $\bB$, and the sign of $J_\parallel$ is positive if the current flows away from star.  The value of $\alpha$ is key to particle acceleration \citep{beloborodov_polar-cap_2008}. As long as $0\leq \alpha <1$ along the magnetic field line, both $\bj$ and $\rho$ can be sustained by a low-energy charge-separated flow extracted from the star by a small $E_\parallel$. Strong particle acceleration and $e^\pm$ creation occurs if $\alpha>1$ or $\alpha<0$. In particular, the current sheet bounding the closed zone has $\alpha<0$. Therefore, it cannot be sustained by particles extracted from the star, and so $e^\pm$ creation is essential in sustaining the current sheet.

\subsection{Full kinetic simulations}

Plasma behavior in the pulsar magnetosphere can be studied from first-principles using global kinetic simulations with explicitly implemented $e^\pm$ creation by accelerated particles. Such simulations were first performed by \citet[hereafter CB14]{chen_electrodynamics_2014}. 
They demonstrated that $e^\pm$ creation is sustained by particle acceleration in the Y-shaped current sheet. Their axisymmetric simulations used a spherical grid in $(\ln r,\theta)$ of size $512\times 512$, which allowed them to choose $\gamma_0=425$.

CB14 observed no $e^\pm$ creation above the polar caps near the axis of the aligned rotator. This is consistent with $0<\alpha<1$ predicted in this region by the FFE model. However, when one takes into account general relativistic corrections, i.e. the Maxwell equations are solved in the curved spacetime around the neutron star, the frame dragging effect can increase $\alpha$ above unity at the poles \citep{beskin_general_1990,muslimov_general_1992,sakai_general_2003}. Global kinetic simulations of \citet{philippov_ab_2015} included this effect, and they observed that it activated pair creation above the polar caps of the aligned rotator.

The axisymmetric pulsar does not pulsate --- the symmetry of the aligned rotator implies that its emission is not modulated by rotation. For this reason alone, it is important to study inclined rotators, which requires full three-dimensional (3D) simulations. Inclined rotators have been studied with global kinetic simulations by \cite{philippov_ab-initio_2014,philippov_ab-initio_2015,cerutti_modelling_2016,Kalapotharakos18}.  At large angles between $\bmu$ and $\bOm$, the magnetosphere significantly changes, as expected from 3D FFE simulations \citep{spitkovsky_time-dependent_2006}. 

In the present paper, we return to the aligned rotator with the goal to better understand its physics using an advanced numerical simulation. We have developed an improved particle-in-cell (PIC) code \href{https://github.com/hoorayphyer/Pigeon}{\texttt{Pigeon}}, with technical advances described in Section~\ref{sec:simulation-setup}. An important change compared with CB14 is the increase of $\gamma_0$ from 425 to $10^4$. This becomes possible because of a much higher grid resolution of $4096\times 4096$. The simulation presented below demonstrates a remarkable concentration of $e^\pm$ creation in the current sheet and provides new insights into the dissipation mechanism, which are discussed in Sections~\ref{sec:struct-magn}-\ref{recon}. 


\section{Simulation setup}
\label{sec:simulation-setup}

The basic setup of our simulation is similar to that in CB14. At time $t=0$ we start with a non-rotating star and a vacuum dipole magnetic field around it. We gradually spin up the star during time $t=5 R_{\star}/c$ to $\Omega=(1/6)(c/R_\star)$, which corresponds to $\RLC=6R_\star$. Then we keep steady rotation. The overall duration of the simulation is $213R_\star/c$, equivalent of $5.7$ revolutions. By this time, a quasi-steady state is established at radii of main interest, $r\simlt 3\RLC$, after an initial relaxation phase.

The stellar surface $r=R_\star$ serves as the inner boundary for the magnetosphere. The star is treated as an ideal conductor with ``frozen'' magnetic field $\bB$. The electric field inside the rotating star is related to the (dipole) magnetic field by the ideal MHD condition $\bE+(\bOm\times\br)\times\bB/c=0$, and the field components $B_r$, $E_\theta$, and $E_\phi$ are continuous across the star's surface. This gives  three boundary conditions, 
\begin{eqnarray} 
B_r(R_\star,\theta) &=& \frac{2\mu\cos\theta}{R_\star^3}, \\
E_\theta(R_\star,\theta)&=&\Omega R_\star \sin\theta\, B_r, \\
E_\phi(R_\star,\theta)&=& 0.
\end{eqnarray}

The simulation domain extends from $R_\star$ to $R_{\rm out}=30R_\star$. At the outer boundary $r=R_{\rm out}$ we implement a damping layer which allows electromagnetic waves (and particles) to escape, with insignificant wave reflection. The large $R_{\rm out}$ is achieved by using the log-spherical grid, with uniform spacing in $\ln r$. The grid has uniform spacing in the polar angle $\theta$.

The emission of photons and their conversion to $e^\pm$ is implemented similarly to CB14, but with a higher threshold for gamma-ray emission, and higher photon energies. Any electron or positron with Lorentz factor $\gamma$ reaching $\gthr=100$ begins to emit photons of energies $\epph=10 m_ec^2$ with rate $\dot{N}_{\pm} = 0.25\gamma c / R_{\star}$. Photons emitted near the star, at radii $r<2R_\star$, convert to $e^\pm$ with a mean free path of $0.2R_\star$, simulating conversion off the magnetic field. Photons emitted at $r>2R_\star=\RLC/3$ convert to $e^\pm$ with a mean free path of $5R_\star$, simulating collision with target soft photons, which are not explicitly included in the simulation. The simulated photon free paths have an exponential distribution around the mean value. The converted photon creates an electron and a positron with equal energies of $\epph/2$. This implementation of pair creation has the main features of the real pair creation process: it is triggered where particles are accelerated to sufficiently high energies, and it occurs with some spatial spreading because of the finite free paths.

A key parameter, which determines the energy scale of the problem is $R_\star\omega_p/c$, where $\omega_p=(4\pi e^2 n/m_e)^{1/2}$ is the  plasma frequency. We define a characteristic  $\omega_p$ using $n=|\rhoco/e|$, where $\rhoco=-\bOm\cdot \bB/2\pi c$ is the corotation charge density. Then,
\beq 
  \omega_p=|2\Omega\omega_B\cos\psi|^{1/2}, \qquad \omega_B=\frac{eB}{m_ec},
\eeq
where $\psi$ is the angle between $\bOm$ and the local $\bB$. For the aligned dipole configuration, $\omega_p$ is highest at the star's poles where $\bB=2\bmu/R_\star^3\parallel\bOm$. The corotation density at the pole defines a characteristic plasma scale,
\beq
  \lambda_p\equiv \frac{c}{\omega_p} \qquad ({\rm pole}).
\eeq

The maximum Lorentz factor of accelerated particles $\gamma_0$ (\Eq~\ref{eq:gam0}) may now be expressed as 
\beq 
\label{eq:g0}
  \gamma_0=\frac{1}{4\,}\frac{R_\star}{\RLC}\,\left(\frac{R_\star}{\lambda_p}\right)^2.
\eeq 
Kinetic simulations are designed to resolve the plasma collective effects, and so the grid for the electromagnetic field must resolve $\lambda_p$. Then, the grid size $N\times N$ determines the maximum achievable $R_\star/\lambda_p$ and thus determines the energy scale $\gamma_0$ according to Equation~(\ref{eq:g0}). It implies that  $\gamma_0$ accessible in the simulation scales as $N^2$. Our grid size of $4096\times 4096$ allows us to increases $\gamma_0$ by a factor of up to $(4096/512)^2$ compared with the CB14 simulation, which used the $512\times 512$ grid. In the simulation presented below we set $\gamma_0=10^4$.

The high $\gamma_0$ is a significant improvement, because it allows one to better separate the relevant scales of the pulsar problem. In particular, there are four important energy scales (in units of $m_ec^2$): $\gamma_0$, $\gthr$, $\epph$, and 1. In a young active pulsar, such as the Crab pulsar, these energies are separated by many orders of magnitude,
\beq 
  \gamma_0\gg\gthr\gg \epph\gg 1,
\eeq 
e.g. $\gamma_0\sim 10^{11}$, $\gthr\sim 10^6$, and $\epph\sim 10^3$. In our simulation, we achieve the separation of the four energy scales by one or two orders of magnitude:
\beq 
   \gamma_0=10^4, \qquad \gthr=100, \qquad \epph=10.
\eeq 
An additional energy scale is set by the ion rest mass $m_ic^2$, which typically satisfies $\epph\simlt m_i/m_e\ll \gthr$. In our simulation we keep the same $m_i/m_e=5$ as in CB14, which is below $\epph$ by a factor of 2. Since the ions do not emit any gamma-rays, $m_i$ should not be important, as long as the ion inertial mass is dominated by kinetic energy, i.e. the ions are accelerated to high Lorentz factors. This regime is satisfied in our simulated aligned rotator: the ions flowing through the magnetosphere to the light cylinder become ultra-relativistic.

Besides the high resolution and good separation of scales, our simulation has a few other improvements. One significant improvement is the inclusion of a thin, dense (gravitationally bound) atmospheric layer on the star's surface. This plasma layer is sustained by injecting low-energy electrons and ions, most of which quickly fall back to the star. The injection rate is regulated so that the atmosphere sustains a desired density $n_{\rm atm}\gg n_{\rm co}$, not depleted by particle lifting  into the magnetosphere. This can be particularly important in the region of high current density, at the footprint of the separatrix current sheet. The atmosphere serves to sustain the electric current demanded by the magnetosphere and to screen magnetospheric electric fields from the star. Then,  particle injection details below the atmosphere are effectively decoupled from the magnetospheric physics.

The atmosphere has an exponential density profile with a scale-height $h$ regulated by the gravitational acceleration. We use simple Newtonian gravity with acceleration $g(r)=g_0(R_\star/r)^2$. Particles are injected with a characteristic thermal speed $v_{\rm th}=0.3c$, and we use a high $g_0=1.8 c^2/R_\star$. We chose the high $v_{\rm th}$ and $g_0$ to reduce the residence time of particles in the atmosphere $v_{\rm th}/g_0$, which helps avoid excessively dense plasma in the layer. The resulting atmosphere is thin, with scale-height $h\approx v_{\rm th}^2/2g_0=R_\star/40$. The density $n_{\rm atm}$ at the base of the atmosphere is maintained at $n\approx 30\,(\Omega B/2\pi c e)$. The local plasma frequency $\omega_p$ that corresponds to $n=n_{\rm atm}$ is resolved by our timestep $\Delta t$ with $\omega_p \Delta t \approx 0.3$.

 Radiation reaction is taken into account in the simulation only for high-energy particles ($\gamma>\gthr$) that emit gamma-rays. Radiation reaction occurs in the form of discrete events of momentum loss: whenever a photon is emitted, it takes momentum from the particle. In addition, for all particles near the star (where the magnetic field is strong and synchrotron cooling should be fast) we implemented damping of gyration. It is implemented by completely removing a particle momentum perpendicular to the magnetic field in the drift frame. The damping is applied where $B>0.1\mu / R_\star^3$. Effectively, in this region, particles are constrained to move along the rotating magnetic field line, like beads on a wire. This prevents artificial magnetic bottles and allows the particles to sink along the field lines toward the star.

We have also implemented $e^\pm$ annihilation, which helps control the number of simulated particles. The minimum particle flux in the equatorial current sheet is $I_0/e$, which would correspond to the positron outflow $\dot{N}_0=2.26\times 10^6\, c/R_\star$ in our simulation. The actual outflow is greater by the multiplicity factor ${\cal M}\gg 1$. We use annihilation at large distances $r>3\RLC$, where the spatial cells $\Delta r\propto r$ become over-populated by the dense $e^\pm$  outflow.


\section{Structure of the magnetosphere}
\label{sec:struct-magn}

The magnetosphere does not find a true steady state and continues to ``breathe," ejecting chunks of plasma (``plasmoids'') along the equatorial plane. However, many of its important features can be studied using time-averaged quantities. The time averaging smears out the plasmoids moving in the equatorial current sheet (which will be discussed separately in Section~\ref{recon}). However, it still gives a sufficiently crispy image of the magnetosphere, and provides a clear picture of dissipation, gamma-ray emission, and $e^\pm$ creation. In this section, we present the magnetospheric structure averaged over two rotation periods during the quasi-steady state observed toward the end of the simulation, between $t=136 R_{\star}/c$ and $213 R_{\star}/c$.

\begin{figure*}[t!]
  \centering
  \includegraphics[width=0.96\textwidth]{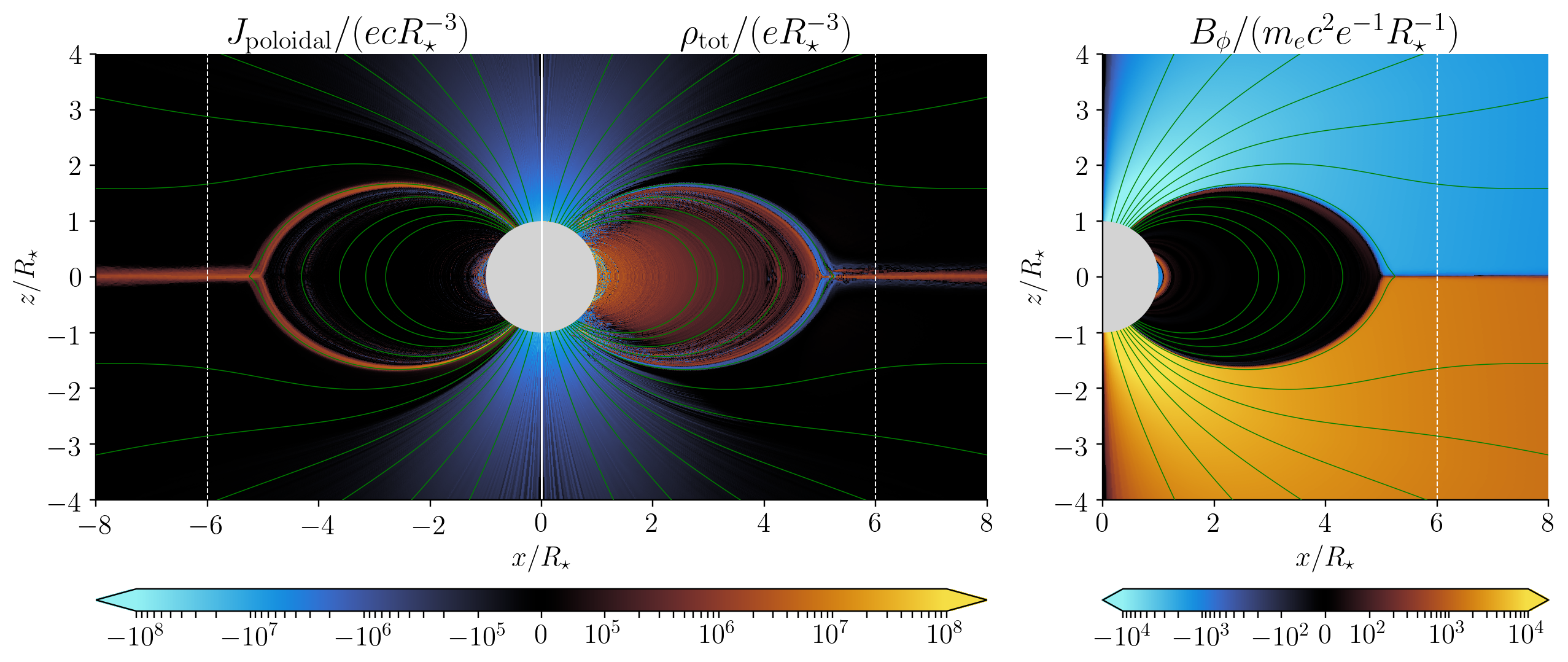}
\caption{Time-averaged poloidal current density $J_{\mathrm{pol}}$, total charge density $\rho$, and toroidal magnetic field $B_{\phi }$. Green curves show the poloidal magnetic field lines (poloidal cross sections of the axisymmetric magnetic flux surfaces), uniformly spaced in the magnetic flux function. The white dashed vertical line indicates the light cylinder. The plots have a resolution of $4^2$ times coarser than the native resolution in the simulation.}
  \label{fig:rhoJB-tave}
\end{figure*}
\begin{figure*}[t!]
  \centering
  \includegraphics[width=0.96\textwidth]{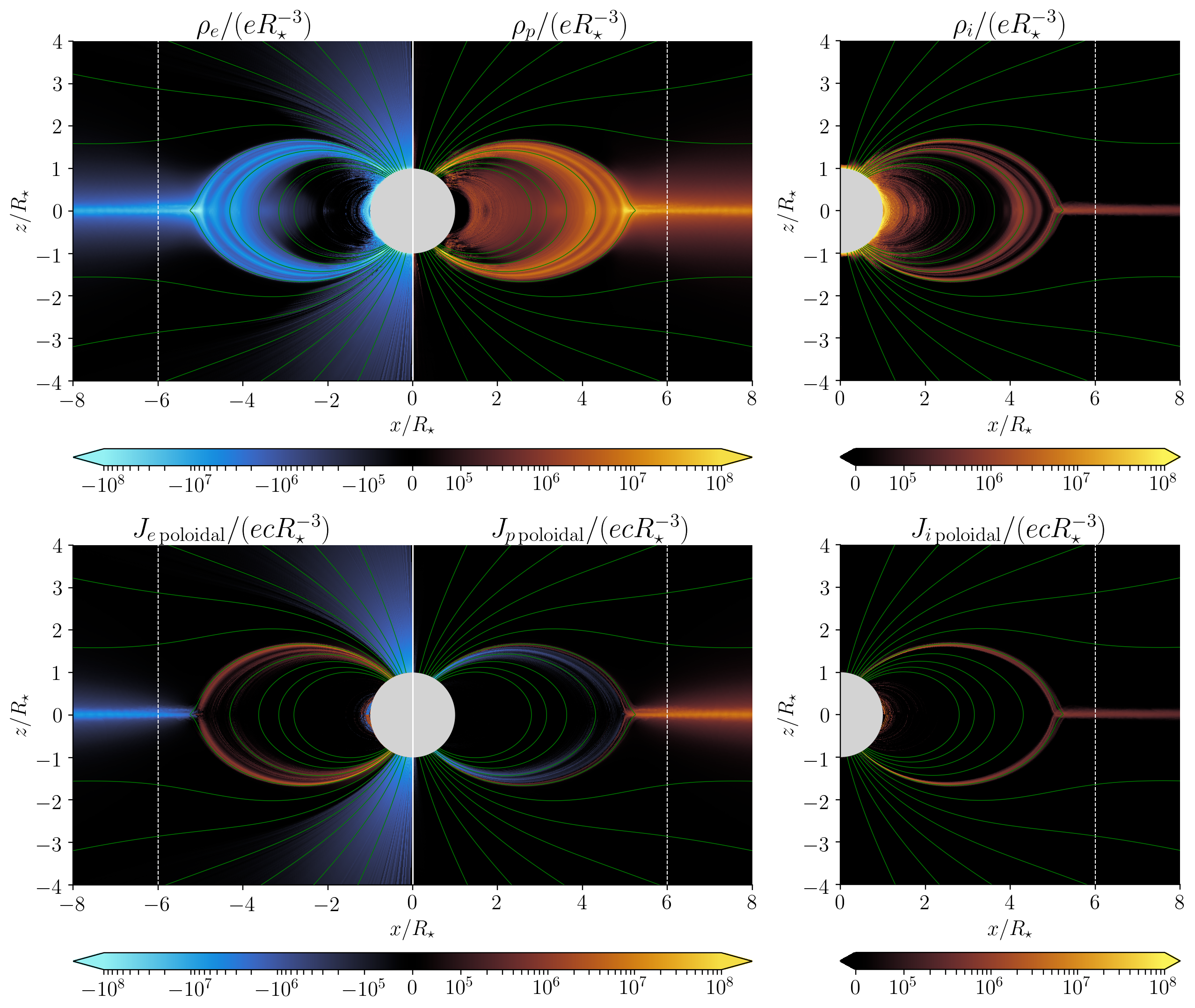}
\caption{Top: time-averaged charge densities of electrons $\rho_e$, positrons $\rho_p$, and ions $\rho_i$. Bottom: contributions of the electrons, positrons, and ions to the poloidal current density. The sign of the poloidal current is set equal to the sign of $J_r$.}
  \label{fig:rhoJepi-tave}
\end{figure*}

Figure~\ref{fig:rhoJB-tave} shows the time-averaged electric current (poloidal component $J_{\rm pol}$), charge density $\rho$, and toroidal magnetic field $B_\phi$. One can see the three basic components predicted by the FFE model: the closed zone with $J_{\rm pol}=0$ and $B_\phi=0$, the negative current from the polar caps (which sustains $B_\phi\neq 0$ in the open field line bundle), and the Y-shaped  current sheet with $J_r>0$ (the return current). The Y-point is located at radius $R_Y<\RLC$.

The observed configuration also displays the charge density $\rho$ predicted by the FFE model. In particular, the separatrix bounding the closed zone is negatively charged \citep{lyubarsky_equilibrium_1990}, and the equatorial current sheet outside the $Y$-point is positively charged. In addition, the kinetic simulation shows that the charged layer along the separatrix at $r<R_Y$ is actually a double layer, resembling a charged capacitor, with a positive surface charge residing just inward of the negatively charged current sheet.

Figure~\ref{fig:rhoJepi-tave} shows the densities and electric currents carried by the three particle species: electrons, positrons, and ions. We observe that the polar-cap current is charge separated, i.e. it is carried by one species --- the electrons extracted from the star. There is practically no $e^\pm$ creation in the polar region. The electrons flow out along the magnetic field lines with modest energies and carry a negligible fraction of the pulsar spindown power. The charge density of the polar outflow is close to the co-rotation density $\rhoco\approx - \bOm\cdot \bB/2\pi c$.

Note that $\rho_{\rm co}$ changes sign along the ``null surface'' where $\bOm\cdot\bB=0$.\footnote{In the dipolar approximation,  $\boldsymbol{\Omega }\cdot \boldsymbol{B} \propto (3 \cos^2\theta - 1) = 0$ is satisfied on the cone $\theta _{\mathrm{null}}\approx 55^\circ$. The actual magnetic field configuration changes from dipolar at $r\sim\RLC$, and the null surface is bent.}
At this surface, the charge-separated outflow was expected to form an ``outer gap'' \citep{cheng_energetic_1986}. We find in our simulation  that this region  is marginally capable of accelerating electrons and positrons to Lorentz factors $\sim \gamma _{\mathrm{thr}}$. Furthermore, the current flowing through this region is small, and therefore the null line does not cause strong dissipation or gamma-ray emission. 
A further increase in resolution and voltage (a higher $\gamma_0/\gthr$) would activate the discharge around the null surface, somewhat helping the ``return'' (positive) current to flow through the region (Bransgrove et al., in preparation).

The positive return current flowing through the magnetosphere of the aligned rotator is dominated by the thin, Y-shaped current sheet. So, to the first approximation, the sheet carries the current 
\beq 
  \Ish\approx I_0,
\eeq 
where $I_0$ is given in Equation~(\ref{eq:I0}). At $r<R_Y$, this current sheet has $\alpha<0$ (\Eq~\ref{eq:alpha}), which prohibits sustaining $\Ish$ by a charge-separated ion flow extracted from the star. Instead, the system employs copious $e^\pm$ creation to sustain $\Ish$. We have measured the time-averaged contributions of electrons, positrons, and ions to this current and found
\begin{equation}
  I_e : I_p : I_i \approx 59\% : 9\% : 32\%.
\end{equation}
Thus, most of $\Ish$ is carried by electrons back-flowing from the Y-point toward the star. This proportion would likely change with increasing multiplicity of $e^\pm$ creation, which may reach huge values, exceeding $10^4$ as suggested by local 1D simulations of $e^\pm$ discharge \citep{Timokhin13}.

As one can see in Figure~\ref{fig:rhoJepi-tave}, the electrons are also flowing outward from the Y-point, along the equatorial plane. Thus, the Y-point acts as a quasi-steady source of electrons (and positrons), which implies a high rate of dissipation and pair creation concentrated at the Y-point. This is an essential feature of the pulsar magnetosphere, which will be discussed in more detail below. 

The high $\gamma_0$ and efficient pair creation allow the separatrix current sheet to become very thin. It is still resolved in the simulation, with $\sim 5$ grid cells across the sheet at its thinnest. We observe that the sheet thickness $\dsh$ is reduced to the minimum possible value,
\beq 
\label{eq:dsh}
  \dsh\approx\frac{\Ish}{ecn},
\eeq
so that the available plasma density in the sheet, $n$, is marginally sufficient to conduct the electric current $\Ish$ with charges moving with relativistic speeds $v\approx c$. This $\dsh$ is explained by the fact that the electromagnetic field dominates the stress-energy tensor and tends to relax as close as possible to the FFE configuration, which has a sharp jump of $B_\phi$ across the separatrix. The steepening of the $B_\phi$ jump stops at $\dsh$ given by \Eq~(\ref{eq:dsh}). At smaller $\dsh$ the current sheet would become charge-starved and fail to conduct $\Ish$ that is needed to support the jump of $B_\phi$.

Everywhere outside the Y-shaped current sheet, the charge density approaches the local co-rotation value $\rhoco$. In particular, it changes sign at the null surface. Interestingly, a large part of the closed zone with $\rhoco>0$ is populated mainly by positrons, rather than ions. The positrons were created by gamma-rays from the equatorial current sheet, and became trapped in the closed zone. Note that the magnetic field lines where positrons dominate have  footprints on the star at $|\theta-90^\circ|>35^\circ$, i.e. in the region of $\rhoco<0$, where {\it electrons} are lifted from the surface rather than the ions (Figure~\ref{fig:rhoJepi-tave}). By contrast, the field lines with footprints  at $|\theta-90^\circ|<35^\circ$ are entirely in the zone of $\rhoco>0$. Here, ions are easily lifted from the surface and suspended in the magnetosphere. On these field lines, positrons have sunk to the star and $\rhoco$ is sustained by the suspended ions. 

\begin{figure*}[t!]
  \centering
  \includegraphics[width=0.96\textwidth]{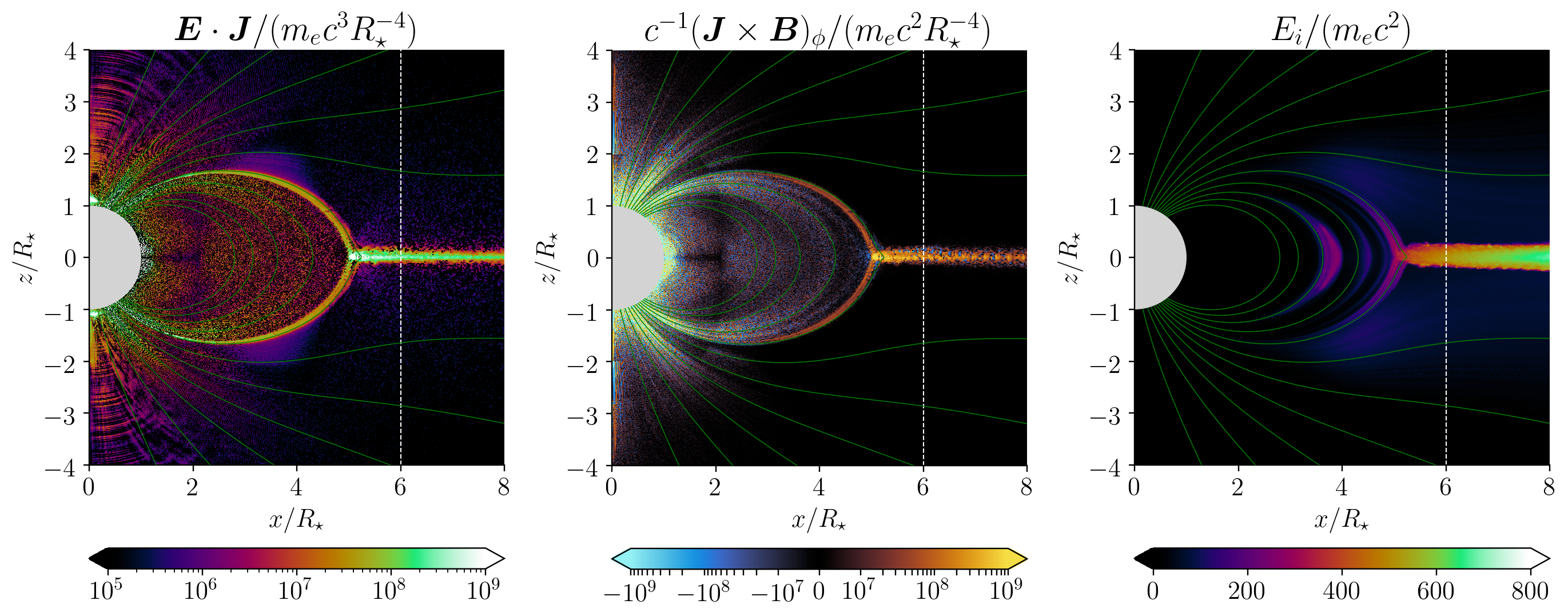}
\caption{Time-averaged $\boldsymbol{E}\cdot \boldsymbol{J}$ (left), $(\boldsymbol{J}\times \boldsymbol{B})_{\phi }/c$ (middle), and the local average of energy in units of $m_ec^2$ (right). 
}
  \label{fig:EdotJJxBEi-tave}
\end{figure*}

The left panel of Figure~\ref{fig:EdotJJxBEi-tave} shows the spatial distribution of the dissipation rate $\bE\cdot\bJ$. The simulation demonstrates that a small fraction of magnetospheric dissipation occurs in the charge-starving separatrix. This dissipation is performed by $E_\parallel$ accelerating particles along the separatrix. The far dominant dissipation occurs in the equatorial current sheet at $r\geq R_Y$. Its most striking feature is the concentration of dissipation in the tiny volume at the Y-point.

\begin{figure*}[t!]
  \centering
  \includegraphics[width=0.96\textwidth]{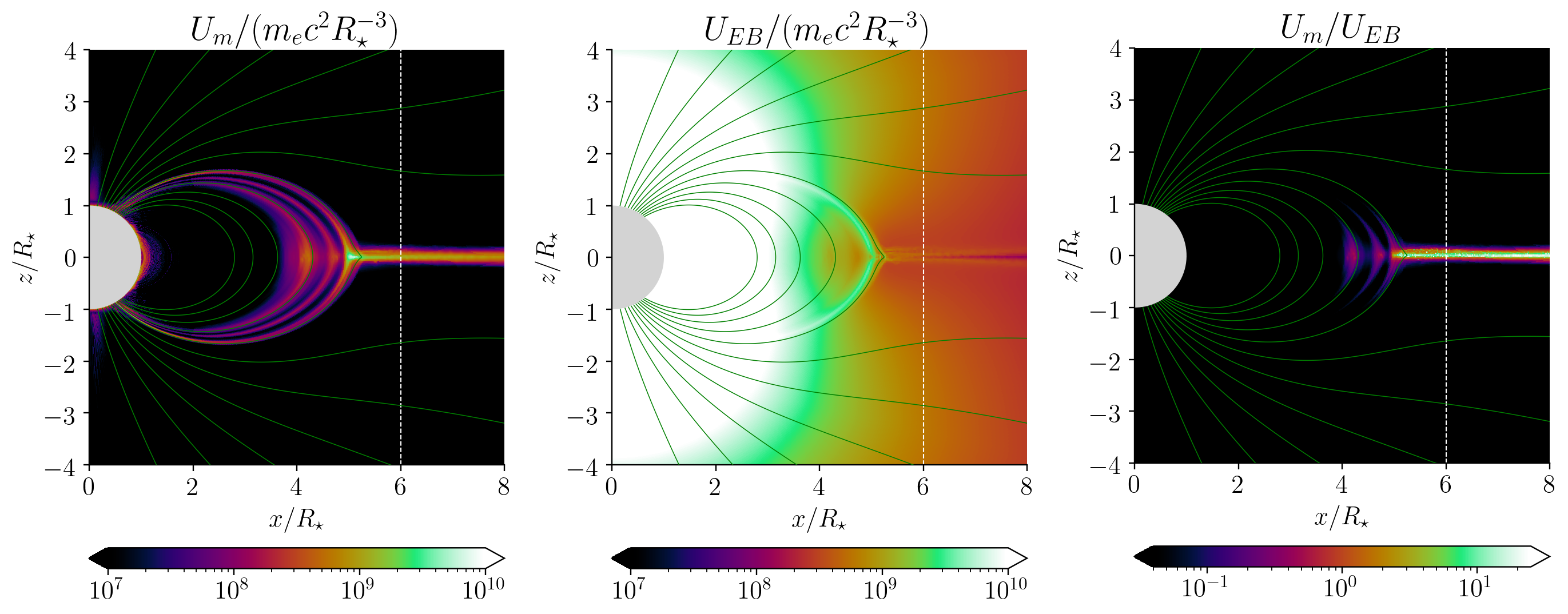}
\caption{Time-averaged matter energy density $U_m$ (left), electromagnetic energy density $U_{EB} = (E^2+B^2)/8\pi$ (middle), and their ratio (right). Note the field enhancement in the separatrix current sheet.
}
  \label{fig:UmUEBUmUEB-tave}
\end{figure*}

The right panel of Figure~\ref{fig:EdotJJxBEi-tave} shows the time-averaged ion energy. The ions have no radiative losses, and their energies serve as a proxy for the accelerating voltage. For instance, the outer gap (across the null surface in the open field line bundle) accelerates ions to $\simlt 100 m_ec^2$. This is comparable to $\gthr m_ec^2$ in our simulation, and hence the outer gap accelerator has a voltage barely capable of triggering pair creation. The ions are accelerated to much higher energies in the equatorial current sheet, up to $\sim 700 m_ec^2$, which is accompanied by a high dissipation rate $\bE\cdot\bJ$ and copious pair creation.

The low ratio $\gthr/\gamma_0=10^{-2}$ offers energetically cheap $e^\pm$ creation, and therefore, one could expect a low dissipation rate in the magnetosphere. We observe that the dissipation rate at $r<R_Y$ is indeed strongly reduced compared with the simulation of CB14, and probably would vanish in the limit of $\gamma_0/\gthr\rightarrow\infty$. However, dissipation in the equatorial current sheet remains quite high. The total dissipation rate measured inside the sphere of radius $2\RLC$ is $L_{\rm diss}\approx 5 \% L_{\rm sd}$. It is $\sim 3$ times lower than previously seen in various simulations of aligned rotators \citep{cerutti_electrodynamics_2017}, however it is still substantial and causes a significant deviation of the observed electromagnetic configuration from the dissipationless FFE model. One visible difference from the configuration found in time-dependent FFE simulations is the position of the Y-point: it is shifted inward from the light cylinder to $R_Y\approx 0.85\RLC$. This shift may persist in the limit of $\gamma_0/\gthr\rightarrow\infty$.\footnote{We have run models with $\gamma_0/\gthr$ 
varying by a factor of 10 and found a similar $R_Y$.}

The energy sink from the electromagnetic field around the Y-point is accompanied by the formation of a heavy $e^\pm$ cloud, which is marginally confined by magnetic stresses, as one can see from Figure~\ref{fig:UmUEBUmUEB-tave}. This is another clear difference from the FFE configuration. The open magnetic field lines serve as an elastic nozzle, which  allows escape of the  plasma from the cloud. As a result, a quasi-steady state is formed, with $e^\pm$ plasma continually accumulated (and intermittently ejected) at the Y-point. 

The dense cloud at the Y-point is a key element of the self-organized magnetosphere, since it feeds the dense electron backflow in the separatrix. This is how the magnetosphere organizes the thin current sheet and almost succeeds in approaching the FFE solution. The FFE configuration is not quite reached, because the backflow density is still limited (resulting in a finite thickness of the current sheet $\dsh$), and more importantly, because sustaining the dense cloud has a significant energy cost.  

\begin{figure}[t] 
  \centering
  \includegraphics[width=0.95\linewidth]{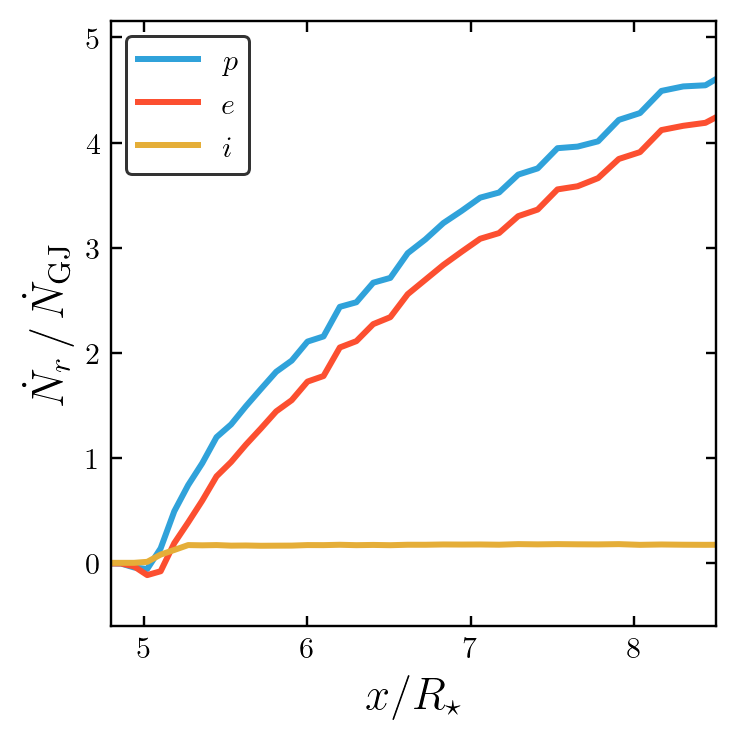}
\caption{The radial flux of particle number (for each of the three species) through a cross section defined by $|z|<0.15R_{\star}$, as a function of $x$ (distance from the rotation axis). The chosen cross section is sufficiently large to cover the equatorial outflow through the magnetic nozzle. Each of the three fluxes was averaged over the last revolution, $176<ct/R_\star<213$, and normalized to $I_0/e=\mu\Omega^2/ce$.
}
  \label{fig:Ndot-r-rec}
\end{figure}

\begin{figure}[t] 
  \centering
  \includegraphics[width=0.95\linewidth]{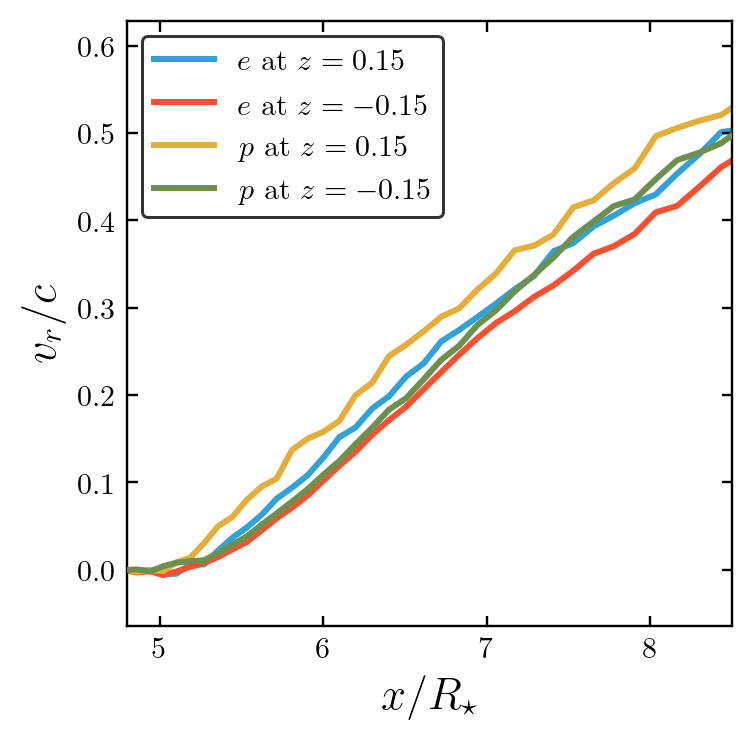}
\caption{Radial hydrodynamic speed $v_{r}$ for each species, measured at $z = \pm 0.15 R_{\star}$, as a function of $x$ (distance from the rotation axis), averaged over the last revolution.
}
  \label{fig:vr-rec}
\end{figure}

The overall pair creation rate $\dot{N}_\pm$ measured in the simulation at $r<1.5\RLC$ gives the $e^\pm$ ``multiplicity'' defined by ${\cal M}\equiv e\dot{N}_\pm/I_0$. We find that its time-averaged value is ${\cal M}\approx 12$. The fast majority of particles flow out in the equatorial plane. Figures~\ref{fig:Ndot-r-rec} and \ref{fig:vr-rec} show the time-averaged radial speed and particle number flux for each species. One can see the steep rise of the electron and positron outflow rate immediately outside the Y-point.



\section{The Super-Rotating Dense Cloud}
\label{Super}

\begin{figure*}[t!]
  \centering
  \includegraphics[width=0.96\textwidth]{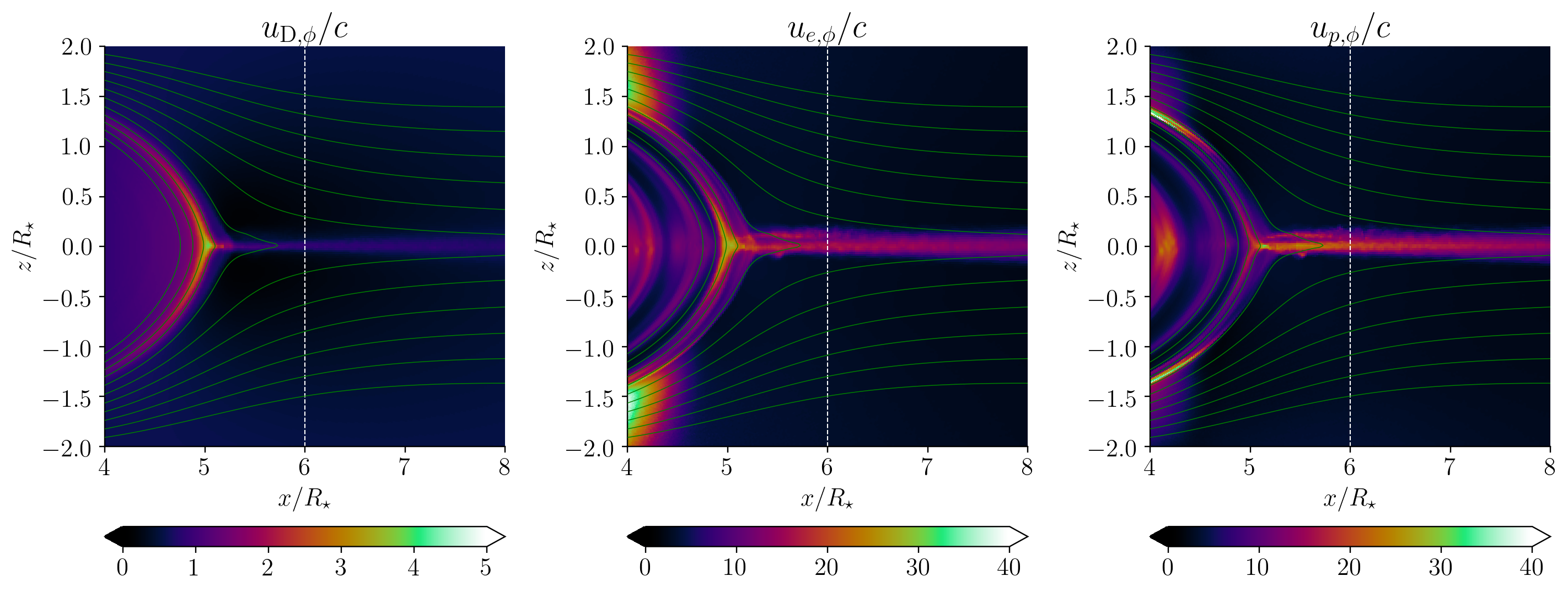}
\caption{Time-averaged toroidal four-velocities $u_{\phi}$ of the electromagnetic field ($\bE\times\bB$ drift, left), electrons (middle), and positrons (right). The figure shows a zoom-in of the Y-point region.}
  \label{fig:uphis-tave}
\end{figure*}

The Y-point cloud has a striking feature: an enormous rotation rate. The co-rotation speed at $R_Y\approx 0.85\RLC$ is $0.85c$, which corresponds to four-velocity $u_{\rm co}\approx 1.6c$. The actual four-velocity of the Y-point plasma observed in the simulation is $u_{e}^{\phi }\sim 30c$ for the electrons and $u_{p}^{\phi }\sim 20c$ for the positrons (Figure~\ref{fig:uphis-tave}). Furthermore, the drift speed of the magnetic field lines near the Y-point corresponds to four-velocity $u_{{\rm D}}^{\phi}\sim 4c$, which exceeds co-rotation,
\beq 
u_{p}^{\phi }, u_{e}^{\phi } \gg u_{{\rm D}}^{\phi}\gg
u_{\rm co}   \qquad \mathrm{(Y~point)}.
\eeq

\subsection{Angular Momentum Flow from the Star }

The high rotation rate at the Y point means that, besides energy dissipation, this region absorbs a lot of angular momentum. Axisymmetry $\partial _{\phi}=0$ implies conservation of the $z$-component of the total angular momentum carried by the plasma and electromagnetic field. This conservation law reads
\begin{align}
  \label{eq:ang_mom}
   \frac{\partial \Laz}{\partial t} + \nabla\cdot \boldsymbol{\Lambda } & = - \be_z\cdot \left[\br\times \left(\rho \bE + \frac{1}{c}\bJ\times \bB\right) \right]  \nonumber \\
   & = - r\sin\theta \left[\rho E_{\phi } + \frac{1}{c}(\bJ\times \bB)_{\phi} \right],
\end{align}
where 
\begin{align}
  \Laz & = r\sin\theta\, \frac{(\boldsymbol{E}\times \bB)_{\phi }}{4\pi c}, \label{eq:Lz} \\
  \boldsymbol{\Lambda  } & = r\sin\theta\, \frac{-E_{\phi }\bE - B_{\phi }\bB + \frac{1}{2}(E^2+B^2)\be_{\phi}}{4\pi},
\label{eq:Lambda}
\end{align}
are the density and flux of angular momentum carried by the electromagnetic field, and $(\be_r,\be_\theta,\be_\phi)$ are the unit vectors of the spherical coordinate system $(r,\theta,\phi)$. 

The angular momentum deposited at the Y-point comes from the rotating star, through the extended region where the electromagnetic field may be approximated as ideal ($\bE\cdot\bB=0$). The flow of angular momentum in an ideal, steady electromagnetic configuration ($\partial _t =0$) admits a simple description. The axisymmetric magnetic flux surfaces are conveniently labeled by the poloidal flux function $f(r,\theta)$, which equals the magnetic flux through the circle of fixed $r,\theta$ ($0<\phi<2\pi$). By design, $f(r,\theta)$ stays constant along the magnetic field lines: $\bB\cdot \nabla f = 0$. Note also that in the steady ideal configuration $\bB\cdot\nabla \Phi=0$, where $\bE=-\nabla\Phi$. Then, one can show that both energy and angular momentum flow along the poloidal flux surfaces,
\begin{equation}
  \label{eq:fluxes}
  \bS\cdot \nabla f = 0, \qquad
  \boldsymbol{\Lambda }\cdot \nabla f = 0,
\end{equation}
where $\bS =c\bE\times \bB/4\pi$ is the Poynting flux, and $\boldsymbol{\Lambda}$ is the angular momentum flux (\Eq~\ref{eq:Lambda}). Hence, the poloidal components of $\boldsymbol{\Lambda}$ and $\bS$ are parallel, and their ratio $\Lambda_{\rm pol}/S_{\rm pol}$ is constant along the flux surfaces. One can then define the specific angular momentum flowing along a flux surface $f$,
\begin{equation}
\label{eq:lz0}
  l(f)\equiv \frac{c^2\Lambda _{\rm pol}}{S_{\rm pol}} = \frac{c^2\Lambda _r}{S_r}= -cr\sin\theta\, \frac{B_r}{E_{\theta }}.
\end{equation}
Its value is set at the stellar surface $r=R_\star$, independently of the shape of the co-rotating magnetosphere. Indeed, both $B_r$ and $E_{\theta }$ are continuous across the stellar surface and related by $cE_{\theta } = - \left[(\bOm\times\br) \times\bB \right]_{\theta}=-\Omega R_{\star}B_r\sin\theta$. This gives a uniform value of $\laz$ across the magnetosphere,
\begin{equation}
  \label{eq:lz}
  \laz = \frac{c^2}{\Omega } = c\RLC.
\end{equation}
This fixed $\laz$ approximately holds throughout the region where dissipation may be neglected, i.e. practically everywhere except the equatorial current sheet. Energy is delivered to the dissipation region along $\bB_{\rm pol}$, and together with energy comes the specific angular momentum $\laz=c\RLC$, which is also damped into the dissipation region.

\subsection{Energy Release and Torque at the Y-point}

Damping of Poynting flux $\bS$ in a steady state is described by  $\nabla\cdot\bS=-\bE\cdot\bJ$. Damping of angular momentum flux $\boldsymbol{\Lambda}$ is performed by the azimuthal component of the Lorenz force $F_\phi=- r\sin\theta \left[\rho E_{\phi } + \frac{1}{c}(\bJ\times \bB)_{\phi} \right]$ (\Eq~\ref{eq:ang_mom}). The Lorentz force $\boldsymbol{F}$ would vanish in FFE, however it is non-zero in the dissipation region, passing the angular momentum from the  electromagnetic field to the plasma.

Neglecting $\rho E_\phi$ (which would completely vanish in a true steady state), we can write 
\beq 
  F_\phi=\frac{|\bJ_{\rm pol}\times\bB_{\rm pol}|}{c}=\frac{J_{\perp} B_{\rm pol}}{c},
\eeq 
where $J_\perp$ is the component of the electric current perpendicular to the magnetic flux surfaces. It is this component of $\bJ$ that creates torque per unit volume,
\beq 
   \dot{\Laz}=r\sin\phi\, F_\phi.
\eeq 
The normalized time-averaged $F_\phi$ is shown in the middle panel of Figure~\ref{fig:EdotJJxBEi-tave}.

Energy is removed from the electromagnetic field and deposited into plasma with rate
\beq
  \dot{U}=\bE\cdot\bJ\approx \bE_{\rm pol}\cdot\bJ_{\rm pol}=E_{\perp}J_{\rm \perp}+E_\parallel J_\parallel. 
\eeq 
We observe in the simulation that energy release at the Y-point is strongly dominated by $E_{\perp}J_{\rm \perp}$. By contrast, energy release in the separatrix at $r<R_Y$ is driven by $E_\parallel$. Thus, there are two qualitatively different mechanisms of particle acceleration: 
\\
(1) $E_\parallel J_\parallel$ accelerates the $\pm$ charges in the opposite directions along the (marginally) charge-starved separatrix. 
\\
(2) $E_\perp J_\perp$ creates torque at the Y-point, which accelerates charges of both signs in the positive $\phi$ direction. Here, the released electromagnetic energy converts to the bulk rotation of the plasma. We observe in the simulation that the localized peak of energy release occurs exactly where $\bJ$ crosses the magnetic flux surfaces, i.e. where a strong $J_\perp$ appears.

\begin{figure}[t!]
  \centering
  \includegraphics[width=0.43\textwidth]{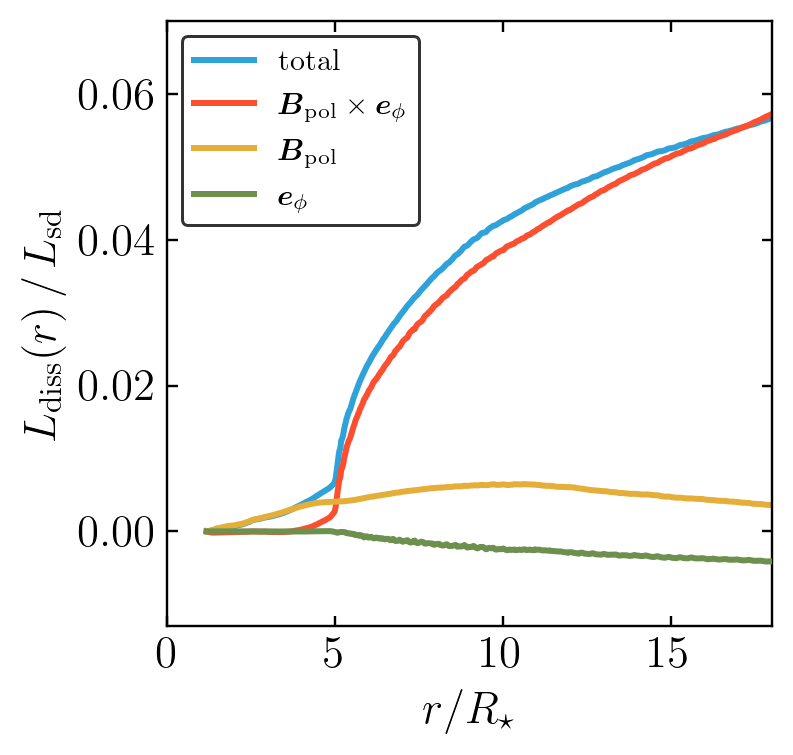}
\caption{Time-averaged electromagnetic power converted to plasma energy inside a sphere of radius $r$, $\Ldiss(r)=\int \bE\cdot\bJ\,dV$, is shown as a function $r$ (blue curve). $\Ldiss(r)$ is normalized to the spindown rate of the star $L_{\rm sd}$. The other three curves show the contributions to $\Ldiss$ from three components of the scalar product  $\bE\cdot\bJ$ in the ``magnetic flux coordinates'' with basis vectors along $\bB_{\rm pol}$, $\be_\phi$, and $ \bB_{\rm pol}\times \be_{\phi}$. The latter basis vector is perpendicular to the magnetic flux surface (parallel to $\nabla f$).}
  \label{fig:EJbreakdown-curve-tave}
\end{figure}

In Section~\ref{recon} we will discuss another dissipation mechanism that works in the equatorial current sheet outside the Y point --- magnetic reconnection.
The overall radial distribution of the time-averaged energy release rate is shown in Figure~\ref{fig:EJbreakdown-curve-tave}.

There is at least one reason for $\bJ$ to flow across the magnetic flux surfaces at the Y-point. Note that the equatorial current beyond the Y-point ($r>R_Y$) flows in the matter-dominated outflow along the magnetic nozzle between the two opposite open magnetic fluxes; thus, it occupies a tiny range of the flux function, $\delta f\approx 0$. The northern and southern parts of the separatrix at $r<R_Y$ each sustains half of the equatorial current, and here the current sheet occupies $\delta f \sim 2\pi r\dsh B_{\rm pol}$ (\Eq~\ref{eq:dsh}). Thus, the return current sheet must somewhat spread across the magnetic flux surfaces as it crosses the Y-point.  

The magnetic torque $\dot{\Laz}$ spins up the plasma at the Y-point to a very high four-velocity $u_\phi$, well beyond $u_{\rm co}$ that corresponds to $v_{\rm co}=\Omega r\sin\theta=\Omega r$. One can try to understand this effect using a toy ``one-zone'' model of a cloud rotating with some velocity $v_\phi$ and experiencing an azimuthal force with uniform density $F_\phi$. The cloud receives energy with rate $v_\phi F_\phi$ and angular momentum with rate $rF_\phi$, and sheds plasma through an outflow with some velocity $v_r$. Then the evolution of the cloud energy density $U$ (which includes rest mass) and angular momentum density $L=rv_\phi U/c^2$ is described by
\beq 
  \frac{dU}{dt}=v_\phi F_\phi - Uv_r, \qquad 
  \frac{dL}{dt}=r F_\phi - Lv_r.
\eeq 
Using these two equations to evaluate $dv_\phi/dt=(c^2/r)d/dt(L/U)$, one finds
\beq 
 \frac{dv_\phi}{dt}=\frac{F_\phi}{U}\left(c^2-v_\phi^2\right).
\eeq 
The cloud is gradually spun up to the asymptotic $v_\phi\rightarrow c$, well beyond $u_{\rm co}$. This toy model may capture the basic reason for the huge rotation rate observed in the simulation. The accumulation of angular momentum may also cause the failure of the system to find a true steady state; instead, the Y-point is observed to ``breathe'' and intermittently eject chunks of plasma.

\begin{figure}[t!]
  \centering
  \includegraphics[width=0.48\textwidth]{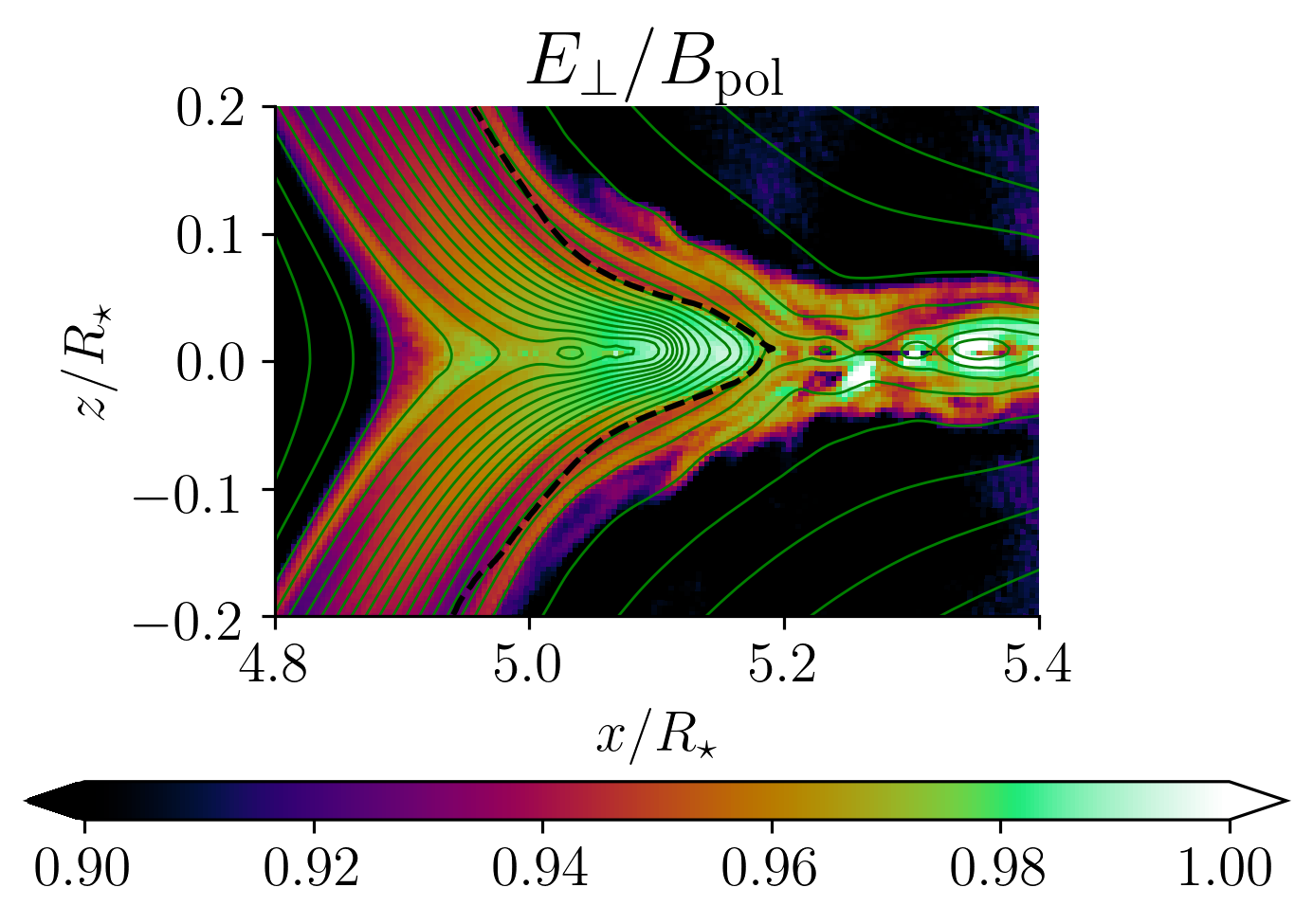}
\caption{Ratio of $E_{\perp}/B_{\mathrm{pol} }$, shown for a snapshot of the simulation at $t = 119 R_{\star}/c$. This figure is a zoom-in of the Y-point region and uses the full resolution of the simulation. Like other figures, green curves show poloidal magnetic field lines (axisymmetric flux surfaces) uniformly spaced in the poloidal  magnetic flux function. The black dashed line is the boundary of the zone of negligible $B_\phi$  ($|B_{\phi }/B_{\mathrm{pol}
    }| < 0.1$). }
  \label{fig:EperpBpol}
\end{figure}

The actual structure of the Y-point differs from the toy one-zone model. The torque is applied to the surface of the dense cloud (the current sheet is thinner than the cloud). The electromagnetic field inside the cloud has $B_\phi=0$ and $E=E_\perp\approx B_{\rm pol}=B$ (Figure~\ref{fig:EperpBpol}), which sustains an ultra-relativistic azimuthal drift speed, $u_{{\rm D}}^{\phi}\sim 4$. The condition $E\approx B$ also suggests that the particles are marginally magnetized, and the cloud may not be well described by the MHD approximation. Furthermore, the cloud rotation is not uniform; the spatial structure of rotation varies as the Y-point breathes and becomes smoothed when viewed with time-averaging. 

Note that the axisymmetric simulation excludes any instabilities in the $\phi$-direction that would be allowed in a 3D simulation. The axisymmetry constraint appears harmless for the Y-point of the aligned rotator, where the closed magnetosphere at $r<R_Y$ creates an impenetrable axisymmetric wall for the return current. In both 2D and 3D models, the return current has to split at $R_Y$ and continue to flow toward the star along the wall. The $\bj\times\bB$ torque accompanying this splitting may be expected even if instabilities break the axisymmetry, however this should be checked with 3D simulations.

\subsection{Gamma-ray Emission and $e^\pm$ Creation}

\begin{figure}[t!]
  \centering
  \includegraphics[width=0.48\textwidth]{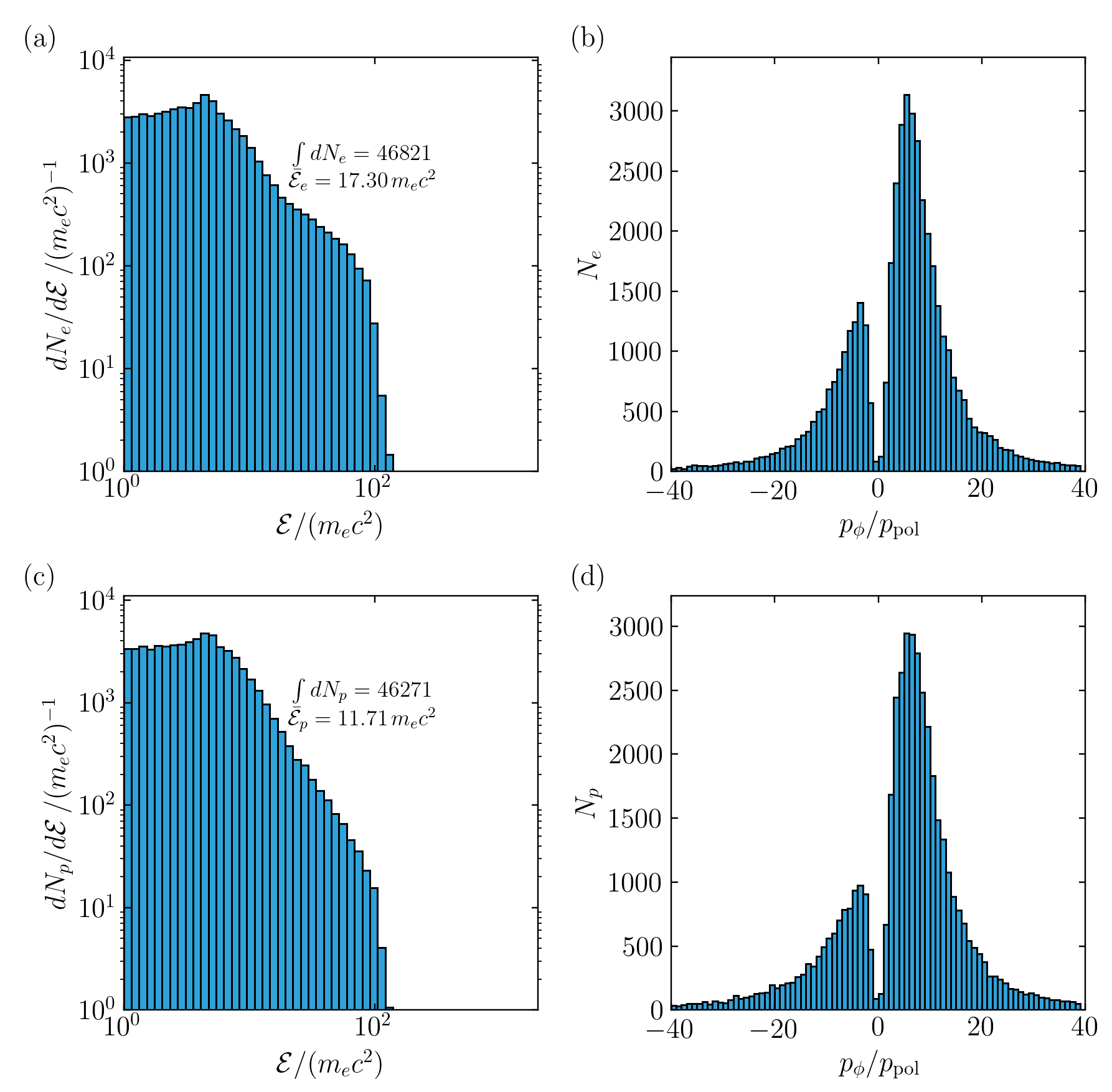}
\caption{Left: energy distribution of electrons (top) and positions (bottom) sampled at $t=98\,R_{\star}/c$ in the zone around the Y-point: $4.9 R_{\star} < x < 5.1 R_{\star}$ and $|z| < 0.6 R_{\star}$. 
Right: distribution of the same sampled particles over $p_\phi/p_{\rm pol}$, the ratio of toroidal and poloidal components of the particle's momentum. One can see that the vast majority of particles in the Y-point region move in the toroidal direction, $|p_\phi/p_{\rm pol}|\gg 1$.}
  \label{fig:dist-Ypoint}
\end{figure}

We observe that the particle distribution function in the Y-point cloud is dominated by the azimuthal motion and has a strong high-energy tail, which extends to $\gthr=100$ (Figure~\ref{fig:dist-Ypoint}). This makes the Y-point a prolific source of gamma-rays and $e^\pm$ pairs. The energetic particles emit gamma-rays along the $\phi$ direction, because the particles themselves have momenta directed along $\phi$, $p_\phi\gg p_{\rm pol}$.

\begin{figure*}[t!]
  \centering
  \includegraphics[width=0.96\textwidth]{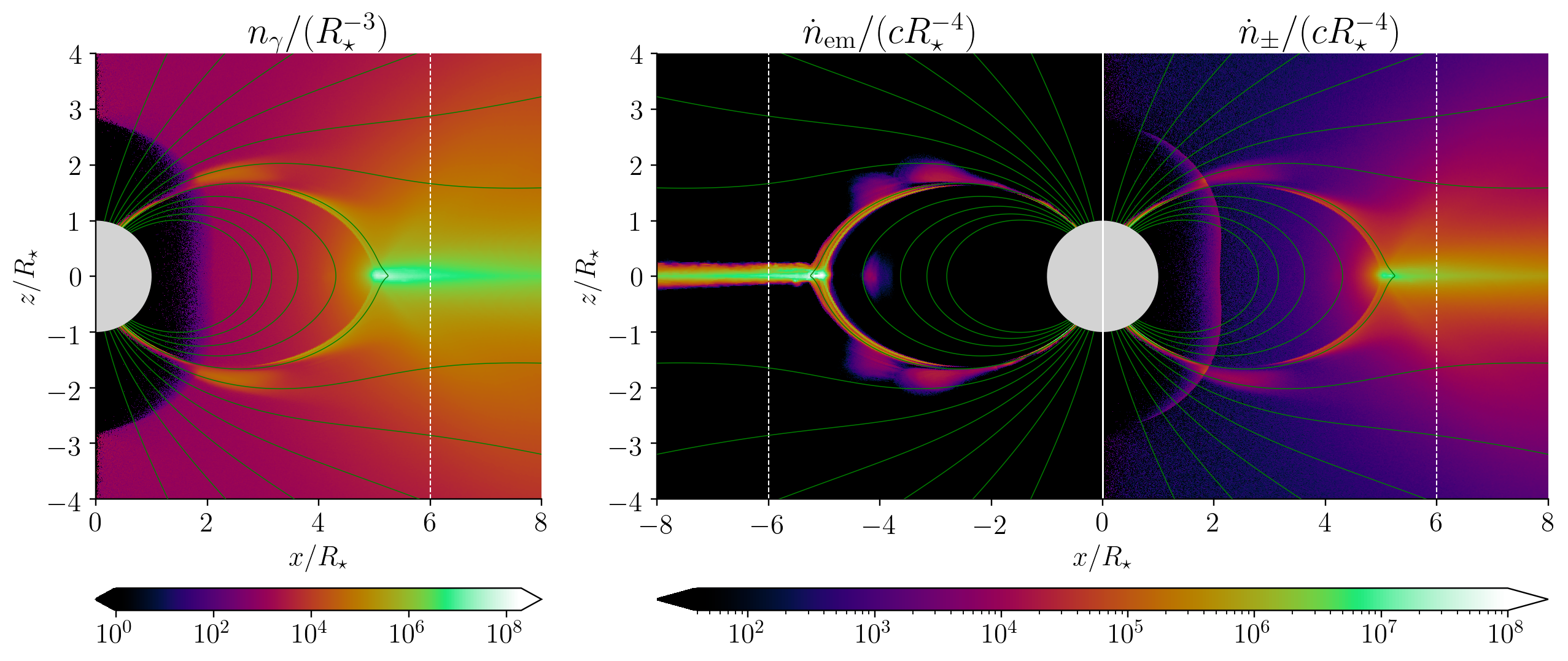}
\caption{Time-averaged  number density of gamma-rays $n_{\gamma }$ (left), gamma-ray emission rate per unit volume $\dot{n}_{\mathrm{em}}$ (middle), and $e^\pm$ creation rate $\dot{n}_{\pm}$ per unit volume (right).}
  \label{fig:nphndotph-tave}
\end{figure*}
\begin{figure}[t!]
  \centering
  \includegraphics[width=0.48\textwidth]{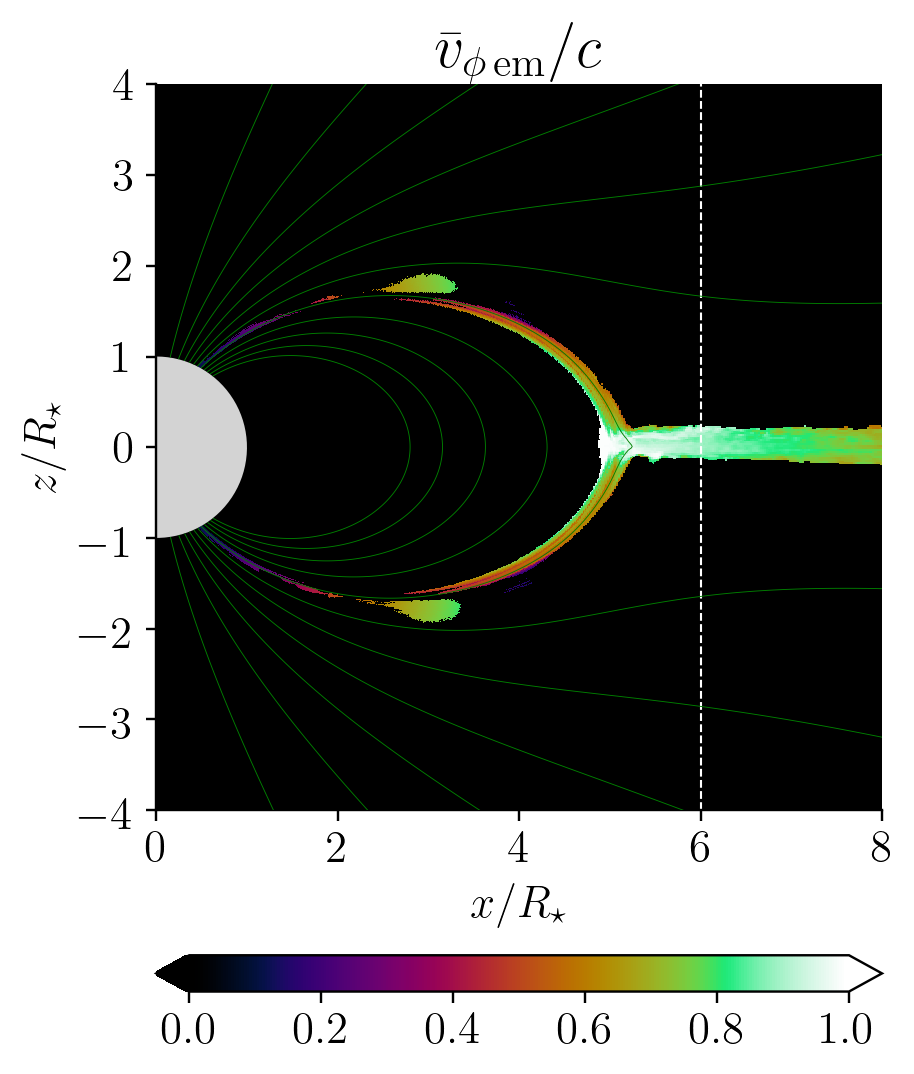}
\caption{Average $\phi $ velocity of an emitted photon.
}
  \label{fig:v3ppem-tave}
\end{figure}

The high plasma density and the concentration of energy release at the Y-point results in the strong concentration of gamma-rays on the circle $r=R_Y$, as seen in the left panel of Figure~\ref{fig:nphndotph-tave}. Figure~\ref{fig:v3ppem-tave} shows the $\phi$ component of the photon velocity at the time of its emission, averaged over many emission events for each spatial cell, $\bar{v}_{\phi \, \mathrm{em}}$. One can see that $\bar{v}_{\phi \, \mathrm{em}}$ nearly equals $c$ at the Y point, i.e. all the photons here are emitted almost exactly in the positive $\phi$ direction. A similar behavior is observed in the current sheet outside the Y-point --  the photon density is also strongly concentrated in the equatorial plane, and the photons are emitted predominantly in the $\phi$ direction.   

The emitted gamma-rays continue to move along straight lines, gradually converting their $\phi$ velocity to a positive radial velocity. Photons convert to $e^\pm$ pairs with the mean free path of $5R_\star$, which implies that pairs are created with a moderate radial displacement $\delta r\sim 2 R_{\star}$ from the photon emission point, with no displacement in $\theta$. As a result the $e^\pm$ creation rate shows a strong concentration in the equatorial plane with a peak near the Y-point (see the right panel in Figure~\ref{fig:nphndotph-tave}).


\section{Magnetic Reconnection in the Equatorial Current Sheet}
\label{recon}

All components of $\bB$ change sign across the equatorial current sheet at $r>R_Y$. The sheet is unstable to the tearing mode, which breaks up the current into threads. This results in reconnection of the opposite open magnetic fluxes, i.e. they partially annihilate each other. Therefore, the  equatorial magnetic nozzle is not ideal --- it is also the site of significant dissipation. 

\begin{figure*}[t]
  \centering
  \includegraphics[width=0.9\linewidth]{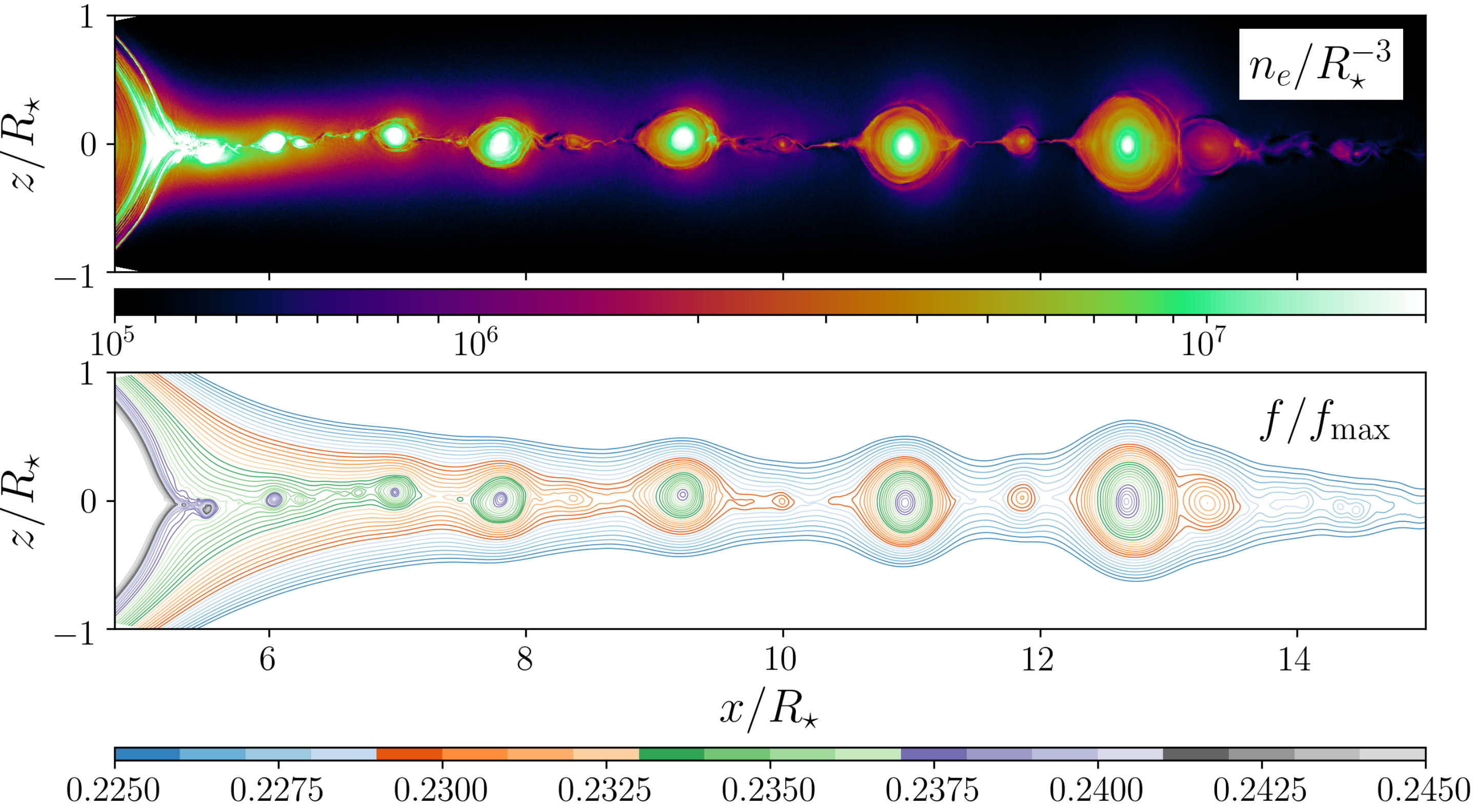}
\caption{Plasmoids formed by magnetic reconnection in the equatorial outflow. The snapshot was taken at $t=213.2\, R_{\star}/c$, at the end of the simulation.
Top: electron density $n_{e}$. Bottom:  contours of the normalized magnetic flux function $f/f_{\max}$.  }
\vspace*{4mm}
  \label{fig:fluxlineflux}
\end{figure*}
\begin{figure}[t]
  \centering
  \includegraphics[width=\linewidth]{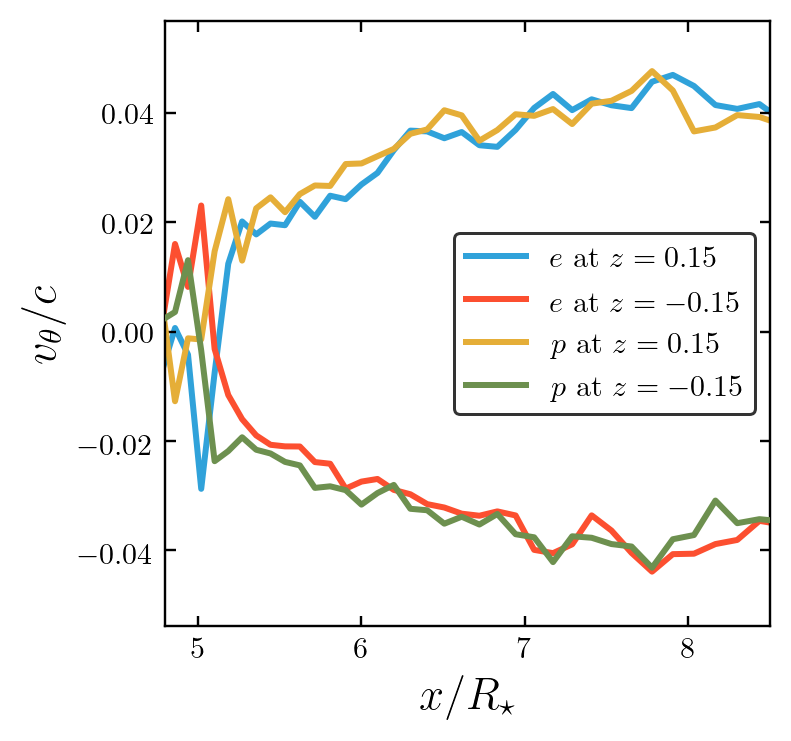}
\caption{Vertical speed of the plasma flow toward the equatorial current sheet. It approximately represents the reconnection speed, which is originally defined as the drift speed $v_D'$ in the wind rest frame $K'$ (where the horizontal drift vanishes). Here, the hydrodynamic speed of electrons was used as a proxy of the plasma flow. It was measured at $z = \pm 0.15 R_{\star}$, as a function of $x$. The values are averaged over the last revolution, $176<ct/R_\star<213$.}
  \label{fig:vth-rec}
\end{figure}

In a full 3D simulation, reconnection would proceed in the plane perpendicular to the current $\bJ$, which has both radial and azimuthal components: $J_r$ sustains the jump of $B_\phi$, and $J_\phi$ sustains the jump of $B_r$. The axisymmetry of our simulation prohibits the tearing instability for $J_r$ (it would develop in the $\phi$ direction in a 3D simulation).  Only tearing of $J_\phi$ develops along $r$, and breaks the current sheet into threads of the azimuthal current. The corresponding reconnection of $B_r$ forms magnetic loops  in the $r$-$\theta$ plane around the threads of $J_\phi$, and the plasma becomes concentrated inside these loops, forming ``plasmoids'' of various sizes. The plasmoids flow away from the star with relativistic speeds. A snapshot of the plasmoid chain in the reconnection layer is shown in Figure~\ref{fig:fluxlineflux}. The chain also sustains the radial component of the equatorial electric current.

The reconnection process involves  advection of horizontal magnetic field $(B_r,B_\phi)$ into the current sheet, and its annihilation. Note that the spiral of open magnetic field lines has $|B_\phi|\sim |B_r|$ at $r\sim \RLC$, and $|B_\phi|\gg |B_r|$ at $r\gg\RLC$. The mechanism of $B_\phi$ dissipation in axisymmetric reconnection may resemble the energy release by $E_\perp J_\perp$ at the Y-point. However, the dissipation driver is different.
Reconnection in the outflow at $r>R_Y$ occurs spontaneously, enabling relaxation of the electromagnetic field toward a state of lower energy. By contrast, the concentrated dissipation at the Y-point is caused by the need for the electric current to connect the separatrix to the equatorial current sheet. 

The spontaneous reconnection in the outflow at $r>R_Y$ is not faithfully reproduced in axisymmetry, as tearing instability in the $\phi$ direction is disabled. Nevertheless, the dissipation rate and the picture of reconnection viewed in the poloidal plane remain very similar to 3D simulations \citep{philippov_ab-initio_2018}. Detailed local 3D simulations of relativistic magnetic reconnection \citep[e.g.][]{sironi_relativistic_2014} also show plasmoid chains similar to the chain observed in Figure~\ref{fig:fluxlineflux}.

Reconnection is usually viewed in the frame where the electromagnetic field drifts directly toward the current sheet, with no parallel velocity component. In our case, this frame is moving horizontally, together with the electromagnetic wind above and below the equatorial plane. The wind is nearly symmetric about the equatorial plane, and its horizontal drift speed is mildly relativistic at radii comparable to $\RLC$. In this frame, the standard reconnection picture would give a vertical inflow and symmetric horizontal outflows in the $\pm r$ directions. The actual reconnection process in pulsars has no such symmetry, because the equatorial current sheet is also the site of a dense, matter-dominated outflow from the Y-point. This energetic outflow results from the huge pair creation rate at $r\approx R_Y$ and flushes outward through the reconnection layer. It may be expected to reduce the reconnection rate near the Y-point, as may be observed in Figure~\ref{fig:vth-rec}. Surprisingly, even far from the Y-point reconnection occurs with about half of the normal rate $v_{\rm rec}\approx 0.1c$  expected for relativistic magnetic reconnection without a guide field (i.e. when all components of $\bB$ change sign across the current sheet).

A detailed view of reconnection is presented in the bottom panel of Figure~\ref{fig:fluxlineflux}, which shows a snapshot with many plotted flux surfaces (lines in the poloidal cross section). They are uniformly spaced in the flux function $f$, so the density of lines reflects the strength of $B_{\rm pol}$. High resolution is helpful for such plots, and we use the native $4096\times 4096$ resolution in the figure. The value of $f$ on the flux surfaces is color coded, which visualizes the ``archeology'' of the closed magnetic loops. One can distinguish two origins of the magnetic loops: ejection from the Y-point and spontaneous magnetic reconnection at larger radii $r>R_Y$. For example, the green field lines at the center of the plasmoid at $x=11.5 R_\star$ came from the Y-point ejection and later were dressed with more magnetic loops by reconnection at $r>R_Y$. Thus, color visualizes the breathing history of the Y-point as well as reconnection at $r>R_Y$.

Practically all large plasmoids observed in the simulation were  initially formed at the Y-point and later grew through spontaneous reconnection at $r>R_Y$, accreting more magnetic flux as they flowed outward. The Y-point ejection approximately spans light green to light orange, $0.245\simlt f/f_{\max}\simlt 0.255$, i.e. the Y-point breaths through $\sim 1$\% of the total magnetic flux of the star $f_{\rm max}$ or $\sim 5\%$ of the open magnetic flux $f_{\rm open}\approx 0.2 f_{\rm max}$. Note also that in the vicinity of the Y-point the equatorial current sheet occupies a tiny $\delta f\sim 10^{-3}f_{\max}$.

One can also see from Figure~\ref{fig:fluxlineflux} that the total magnetic flux entering the dissipative equatorial current sheet spans the range $0.235\simlt f/f_{\rm max}\simlt 0.255$. Thus, the total flux $\delta f\approx 0.1f_{\rm open}$ penetrates the equatorial current sheet. This sets the total energy budget for the equatorial dissipation. The energy is delivered from the star to the dissipation region along the flux surfaces, and overall $\sim 6\%$ of the spindown power $L_{\rm sd}$ becomes dissipated. The Poynting flux along flux surfaces $0.245\simlt f\simlt 0.255$ mainly goes to the Y-point dissipation, and the Poynting flux along flux surfaces $0.235\simlt f\simlt 0.245$ is released through reconnection in the outer equatorial current sheet. Part of the released energy is carried by the outflowing plasma and part by the escaping gamma-rays.


\section{Conclusions}

This paper has presented a global relativistic kinetic simulation of an axisymmetric pulsar magnetosphere with self-consistent $e^\pm$ pair production. Our PIC simulation used log-spherical coordinates with a grid size of $4096\times 4096$ covering the radial domain $r<30R_\star$, several times larger than the pulsar light cylinder chosen in the model, $\RLC=6 R_\star$. The simulation has advanced features including a thin, dense atmospheric layer sustained on the star's surface, which provides a sufficient reservoir of particles at the magnetospheric footprints. Importantly, the simulation did not impose any plasma injection into the magnetosphere. Pairs were created only in response to particle acceleration, through a two-step process --- emission of gamma-rays followed by their conversion to $e^\pm$, as occurs in real pulsars. The high resolution allowed us to push the maximum accelerating voltage to a high value, which corresponds to the electron  Lorentz factor $\gamma_0=10^4$, and to achieve a good separation of important energy scales $\gamma _0:\gamma _{\mathrm{thr}}: \varepsilon _{\mathrm{ph}} = 10^4:10^2:10$. 

Such first-principles numerical experiments aim to uncover physical mechanisms operating in pulsars, in particular how the plasma magnetosphere self-organizes through pair creation and dissipates electromagnetic energy. As a disclaimer, we note that the numerical experiments do not reconstruct concrete observed pulsars. The achieved separation of scales and the multiplicity of $e^\pm$ creation are still below those in real objects.
In addition, photon free paths were drawn from a simple prescribed distribution. More detailed calculations of photon-photon collisions will require expensive simulations of radiative transfer (and vastly vary between different pulsars).

Our main results may be summarized as follows.

The energy release and $e^\pm$ creation are strongly concentrated in the thin, Y-shaped current sheet, with a peak in a small volume at the Y-point. This is a remarkable feature, especially taking into account that the mean free-path of gamma-rays near the light cylinder
is comparable to $\RLC$ in our simulation. This self-organized concentration of $e^\pm$ creation is achieved by the system via developing an enormous rotation rate at the Y-point, which results in nearly perfect beaming of gamma-ray emission along the azimuthal direction. The plasma develops super-rotation by absorbing the angular momentum flux flowing from the star along the poloidal magnetic lines.

The simulation shows that the radius of the Y-point circle $R_Y$ is shifted inward from the light cylinder by about $15\%$, and ``breathes'' with a small amplitude around this average position. This implies a mildly relativistic co-rotation speed at the Y-point, $v_{\rm co}\approx 0.85c$. The actual ultra-relativistic rotation of the plasma far exceeds co-rotation with the star. Therefore, we call it super-rotation.

We have studied in detail the Y-shaped current sheet. The separatrix current at $r<R_Y$ is mainly supported by the electron backflow from the dense Y-point cloud (with smaller fractions supplied by $e^\pm$ discharge in the separatrix itself and by ion extraction from the star). The thickness of the separatrix current sheet $\dsh$ is self-regulated to marginal charge starvation and therefore related to the plasma density at the Y-point. The system achieves a small $\dsh$ by sustaining a high density at $R_Y$ via the concentrated pair creation. The accumulated pair density at the Y-point is limited by the confinement condition, as plasma excess is intermittently ejected into the equatorial outflow through the elastic magnetic nozzle at $R_Y$. It will be interesting to investigate in future simulations whether a higher multiplicity of $e^\pm$ creation near the star could significantly change the processes at the light cylinder. Simulations by \cite{Chen20} suggest that a plasma outflow from $e^\pm$ discharge near the star may become sufficient to form the Y-shaped current sheet and sustain the return current. We expect that the strong torque and dissipation near the Y-point will persist in any case, and the separatrix will always self-regulate to marginal charge starvation.

The equatorial current sheet at $r>R_Y$ also tends to become thin (close to charge starvation) between the massive plasmoids that carry the ejected plasma. In contrast to the separatrix at $r<R_Y$, the equatorial current sheet is unstable to tearing, and releases significant power through magnetic reconnection.

Our simulation shows that for pulsars far from the death line (i.e. in the limit of $\gamma_0\gg \gthr$) only a small fraction of dissipation occurs in the separatrix at $r<R_Y$, through  particle acceleration by $E_\parallel$ along the magnetic field lines. Far greater energy release occurs in the equatorial plane at $r\geq R_Y$, especially near the Y-point. Its mechanism is driven by $E_\perp$ and qualitatively differs from particle acceleration by $E_\parallel$. In particular, the concentrated peak of energy release at the Y-point occurs through spinning up the plasma to super-rotation, followed by its intermittent ejection through the nozzle between the two open magnetic fluxes.

The beaming of the produced gamma-rays along the azimuthal direction, with the peak of emission tangential to the circle $R_Y$, may have important observational implications, in particular if this feature persists in 3D simulations of inclined rotators. Magnetospheres of inclined rotators significantly differ from the aligned rotator studied in this paper. More work on inclined rotators with various magnetic fields and rotation rates is needed before first-principle simulations help interpret the diverse spectra and pulse profiles of observed objects. We also emphasize that our simulation is relevant only to pulsars capable of pair creation at the light cylinder.

\medskip

The simulation was performed using computational resources provided by the NASA High-End Computing (HEC) Program through the NASA Advanced Supercomputing (NAS) Division at Ames Research Center.
A.M.B. acknowledges support by NASA grant NNX 17AK37G, NSF grant AST 2009453, Simons Foundation grant \#446228, and the Humboldt Foundation.

\bibliographystyle{aasjournal.bst}
\bibliography{ms}

\end{document}